\documentclass[aps,prd,superscriptaddress,twoside,twocolumn,nofootinbib,showpacs,floatfix]{revtex4-1}
\usepackage{amsmath,amssymb}
\usepackage{graphicx}
\usepackage{subfigure}
\usepackage[colorlinks]{hyperref}
\usepackage[usenames,dvipsnames]{color}
\pdfpagewidth=\paperwidth
\pdfpageheight=\paperheight
\allowdisplaybreaks
\usepackage{braket}

\begin{document}

\title{\boldmath Analysis of the $\gamma n \to K^+ \Sigma^-(1385)$ photoproduction}

\author{Ai-Chao Wang}
\affiliation{School of Nuclear Science and Technology, University of Chinese Academy of Sciences, Beijing 101408, China}

\author{Neng-Chang Wei}
\affiliation{School of Nuclear Science and Technology, University of Chinese Academy of Sciences, Beijing 101408, China}

\author{Fei Huang}
\email[Corresponding author. Email: ]{huangfei@ucas.ac.cn}
\affiliation{School of Nuclear Science and Technology, University of Chinese Academy of Sciences, Beijing 101408, China}

\date{\today}

\begin{abstract}
In our previous work [Phys. Rev. D $\bf{101}$, 074025 (2020)], the photoproduction $\gamma p \to K^+ \Sigma^0(1385)$ has been investigated within an effective Lagrangian approach. There, the reaction amplitudes were constructed by considering the $t$-channel $K$ and $K^\ast(892)$ exchanges, $s$-channel $N$ contribution, $u$-channel $\Lambda$ exchange, generalized contact term, and a minimum number of $s$-channel $N$ and $\Delta$ resonance diagrams. It was found that the inclusion of one of the $N(1895){1/2}^-$, $\Delta(1900){1/2}^-$, and $\Delta(1930){5/2}^-$ resonances is essential to reproduce the available differential and total cross-section data for $\gamma p \to K^+ \Sigma^0(1385)$. In the present work, we employ the same model to study the photoproduction $\gamma n \to K^+ \Sigma^-(1385)$, with the purpose being to understand the reaction mechanism and, in particular, to figure out which one of the $N(1895){1/2}^-$, $\Delta(1900){1/2}^-$, and $\Delta(1930){5/2}^-$ resonances is really capable for a simultaneous description of the data for both $K^+ \Sigma^0(1385)$ and $K^+ \Sigma^-(1385)$ photoproduction reactions. The results show that the available data on differential and total cross sections and photo-beam asymmetries for $\gamma n \to K^+ \Sigma^-(1385)$ can be reproduced only with the inclusion of the $\Delta(1930){5/2}^-$ resonance rather than the other two. The generalized contact term and the $t$-channel $K$ exchange are found to dominate the background contributions. The resonance $\Delta(1930){5/2}^-$ provides the most important contributions in the whole energy region considered, and it is responsible for the bump structure exhibited in the total cross sections.
\end{abstract}



\maketitle

\section{Introduction}   \label{Sec:intro}

The study of nucleon and $\Delta$ resonances ($N^\ast$s and $\Delta^\ast$s) has been of great interest in hadron physics, since a deeper understanding of $N^\ast$s and $\Delta^\ast$s can help us get insight into the nonperturbative regime of quantum chromodynamics (QCD). It is well known that most of our current knowledge about $N^\ast$s and $\Delta^\ast$s is coming from $\pi N$ scattering or $\pi$ photoproduction reactions. Nevertheless, there are lots of resonances predicted by quark model calculations \cite{Isgur:1977ef, Koniuk:1979vy} or lattice QCD simulations \cite{Edwards:2013, Lang:2017, Kiratidis:2017, Andersen:2018} but not detected in the $\pi N$ production experiments. This situation forces us to study $N^\ast$s and $\Delta^\ast$s in other reaction channels, to which some of the $N^\ast$s and $\Delta^\ast$s might have much stronger coupling strengths than to the $\pi N$ channel. In this regard, the production processes of heavier mesons such as $\eta$, $\eta' $, $K$, $K^\ast$, $\omega$, and $\phi$ provide us rather effective tools to investigate the high-mass resonances, as these channels have much higher threshold energies than that of $\pi N$. In the present work, we concentrate on the photoproduction reaction of $\gamma n \to K^+ \Sigma^-(1385)$.

Experimentally, the high-statistics data on differential cross sections and photo-beam asymmetries for the reaction $\gamma n \to K^+ \Sigma^-(1385)$ became available in recent years \cite{Paul:2014,Hicks:2009}. In 2009, the LEPS Collaboration released the first high-statistics differential cross-section data in the photon laboratory incident energy range $E_{\gamma}=1.5-2.4$ GeV and the first photo-beam asymmetry data at three energy points \cite{Hicks:2009}. In 2014, the CLAS Collaboration at the Thomas Jefferson National Accelerator Facility (JLab) reported the preliminary differential cross-section data in the photon laboratory incident energy range $E_{\gamma}=1.6-2.4$ GeV \cite{Paul:2014}. The CLAS data are located in a larger scattering angle range than the LEPS data.

Theoretically, three works based on effective Lagrangian approaches have already been devoted to the study of $\gamma n \to K^+ \Sigma^-(1385)$ reaction \cite{Zou:2010,Xiaoyun:2016,Byung:2017}. In Ref.~\cite{Zou:2010}, the LEPS differential and total cross-section data and beam asymmetry data for $\gamma n \to K^+ \Sigma^-(1385)$ together with the CLAS total cross-section data for $\gamma p \to K^+ \Sigma^0(1385)$ were analyzed, and it was reported that a newly proposed $\Sigma$ $(J^p={1/2}^-)$ state with a mass around $1380$ MeV is helpful to reproduce the negative beam asymmetry data for $\gamma n \to K^+ \Sigma^-(1385)$. We mention that at that time the high-precision differential cross-section data for $\gamma p \to K^+ \Sigma^0(1385)$ from the CLAS Collaboration \cite{Moriya2013} were not yet being available, thus they were not included in the analysis of Ref.~\cite{Zou:2010} to constrain their model. In Ref.~\cite{Xiaoyun:2016}, the cross-section data for $\gamma n \to K^+ \Sigma^-(1385)$ from both the LEPS and CLAS Collaborations were analyzed within an interpolated Regge model, and it was reported that the data can be described without inclusion of any $s$-channel baryon resonances. Note that the LEPS photo-beam asymmetry data for $\gamma n \to K^+ \Sigma^-(1385)$ were not considered in Ref.~\cite{Xiaoyun:2016}, and it is also not clear whether this model can describe the CLAS high-precision differential cross-section data for $\gamma p \to K^+ \Sigma^0(1385)$ or not. In Ref.~\cite{Byung:2017}, the available data on differential cross sections for both $\gamma p \to K^+ \Sigma^0(1385)$ and $\gamma n \to K^+ \Sigma^-(1385)$ and on beam asymmetries for $\gamma n \to K^+ \Sigma^-(1385)$ were investigated in a Regge model. The data were qualitatively described with some room left for improvements as no resonance was considered in this model.

In our previous work \cite{Wang:2020}, we have performed a detailed investigation of the $\gamma p \to K^+ \Sigma^0(1385)$ reaction within an effective Lagrangian approach.\footnote{As mentioned in Ref.~\cite{Wang:2020}, the $\gamma p \to K^+ \Sigma^0(1385)$ reaction has also been investigated in Refs.~\cite{Yong:2008,hejun:2014} in effective Lagrangian approaches.} We considered the $s$-channel $N$ contribution, $t$-channel $K$ and $K^\ast$ exchange, $u$-channel $\Lambda$ exchange, and generalized contact term. Besides, the contributions from a minimum number of $N$ and $\Delta$ resonances in the $s$ channel was also introduced in constructing the reaction amplitudes. The CLAS high-precision data on differential cross sections were well described, and it was found that one of the $N(1895){1/2}^-$, $\Delta(1900){1/2}^-$, and $\Delta(1930){5/2}^-$ resonances was necessarily needed to reproduce the data.

In the present work, we employ the same model as in our earlier work of Ref.~\cite{Wang:2020} that works quite well for the $\gamma p \to K^+ \Sigma^0(1385)$ reaction to study the photoproduction reaction $\gamma n \to K^+ \Sigma^-(1385)$. The purpose is to get a unified description of both $\gamma p \to K^+ \Sigma^0(1385)$ and $\gamma n \to K^+ \Sigma^-(1385)$  within the same theoretical model. Particularly, we want to figure it out which one of the $N(1895){1/2}^-$, $\Delta(1900){1/2}^-$, and $\Delta(1930){5/2}^-$ resonances can results in a simultaneous description of the data for both $K^+ \Sigma^0(1385)$ and $K^+ \Sigma^-(1385)$ photoproduction reactions.

The paper is organized as follows. In Sec.~\ref{Sec:formalism}, we briefly introduce the framework of our theoretical model. In Sec.~\ref{Sec:results}, we present our theoretical results and discussions. Finally, a brief summary and conclusions are given in Sec.~\ref{sec:summary}.

\section{Formalism}  \label{Sec:formalism}

\begin{figure}[tbp]
\subfigure[~$s$ channel]{
\includegraphics[width=0.45\columnwidth]{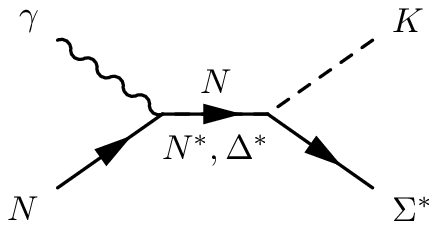}}  {\hglue 0.4cm}
\subfigure[~$t$ channel]{
\includegraphics[width=0.45\columnwidth]{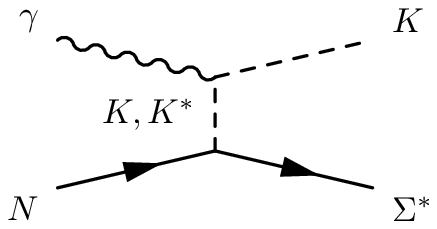}} \\[6pt]
\subfigure[~$u$ channel]{
\includegraphics[width=0.45\columnwidth]{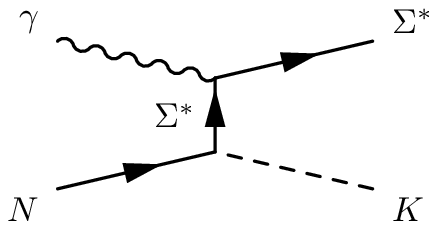}} {\hglue 0.4cm}
\subfigure[~Interaction current]{
\includegraphics[width=0.45\columnwidth]{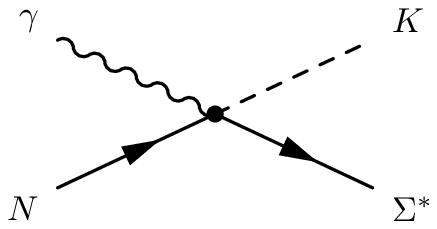}}
\caption{Generic structure of the amplitude for $\gamma n\to K^+ \Sigma^-(1385)$. Time proceeds from left to right. The symbols $\Sigma^\ast$ and $K^\ast$ denote $\Sigma(1385)$ and $K^\ast(892)$, respectively.}
\label{FIG:feymans}
\end{figure}

The full photoproduction amplitudes for $\gamma N \to K \Sigma(1385)$ can be expressed as \cite{Haberzettl:1997,Haberzettl:2006}
\begin{equation}
M^{\nu\mu} = M^{\nu\mu}_s + M^{\nu\mu}_t + M^{\nu\mu}_u + M^{\nu\mu}_{\rm int},  \label{eq:amplitude}
\end{equation}
with $\nu$ and $\mu$ being the Lorentz indices of $\Sigma(1385)$ and the photon $\gamma$, respectively. The first three terms $M^{\nu\mu}_s$, $M^{\nu\mu}_t$, and $M^{\nu\mu}_u$ stand for the $s$-, $t$-, and $u$-channel pole diagrams, respectively, with $s$, $t$, and $u$ being the Mandelstam variables of the internally exchanged particles. They arise from the photon attaching to the external particles in the $KN\Sigma(1385)$ interaction vertex. The last term, $M^{\nu\mu}_{\rm int}$, stands for the interaction current that arises from the photon attaching to the internal structure of the $KN\Sigma(1385)$ interaction vertex. All four terms in Eq.~(\ref{eq:amplitude}) are diagrammatically depicted in Fig.~\ref{FIG:feymans}.

The interaction mechanisms considered in the present work for $\gamma n \to K^+ \Sigma^-(1385)$ are the same as those in Ref.~\cite{Wang:2020} that works for $\gamma p \to K^+ \Sigma^0(1385)$, except that the $u$-channel $\Lambda$ exchange in $\gamma p \to K^+ \Sigma^0(1385)$ is now replaced by the $\Sigma$ exchange in $\gamma n \to K^+ \Sigma^-(1385)$ due to the charges of $\Sigma(1385)$ baryons. In brief, the following contributions (as shown in Fig.~\ref{FIG:feymans}) are considered in constructing the $s$-, $t$-, and $u$-channel amplitudes for $\gamma n \to K^+ \Sigma^-(1385)$: (i) $N$, $N^\ast$, and $\Delta^\ast$ contributions in the $s$ channel, (ii) $K$ and $K^\ast(892)$ exchanges in the $t$ channel, and (iii) $\Sigma^\ast$ hyperon exchange in the $u$ channel.

Using an effective Lagrangian approach, one can, in principle, obtain explicit expressions for these amplitudes by evaluating the corresponding Feynman diagrams. However, the exact calculation of the interaction current $M^{\nu\mu}_{\rm int}$ is impractical, as it obeys a highly nonlinear equation and contains diagrams with very complicated interaction dynamics. Following Refs.~\cite{Haberzettl:1997,Haberzettl:2006,Huang:2012,Huang:2013}, we model the interaction current by a generalized contact current, which effectively accounts for the interaction current arising from the unknown parts of the underlying microscopic model:
\begin{equation}
M^{\nu\mu}_{\rm int} = \Gamma^\nu_{\Sigma^\ast N K}(q) C^\mu + M^{\nu\mu}_{\rm KR} f_t.  \label{eq:Mint}
\end{equation}
Here $\nu$ and $\mu$ are Lorentz indices for $\Sigma(1385)$ and $\gamma$, respectively; $\Gamma^\nu_{\Sigma^\ast NK}(q)$ is the vertex function of $\Sigma(1385) NK$ coupling,
\begin{equation}
\Gamma^\nu_{\Sigma^\ast NK}(q) = - \frac{g_{\Sigma^\ast NK}}{M_K} q^\nu,
\end{equation}
with $q$ being the four-momentum of the outgoing $K$ meson; $M_{\rm KR}^{\nu\mu}$ is the Kroll-Ruderman term,
\begin{equation}
M^{\nu\mu}_{\rm KR} = \frac{g_{\Sigma^\ast NK}}{M_K}  g^{\nu\mu} TQ_K,
\end{equation}
with $T$ denoting the isospin factor of the $\Sigma(1385) NK$ coupling and $Q_K$ being the electric charge of the outgoing $K$ meson; $f_t$ is the phenomenological form factor attached to the amplitude of $t$-channel $K$ exchange, whose explicit form has been given in Ref.~\cite{Wang:2020}; $C^\mu$ is an auxiliary current, which is nonsingular and is introduced to ensure that the full photoproduction amplitudes of Eq.~(\ref{eq:amplitude}) are fully gauge invariant. Following Refs.~\cite{Haberzettl:2006,Huang:2012}, we choose $C^\mu$ for $\gamma n \to K^+ \Sigma^{-}(1385)$ as
\begin{align}
C^\mu = & - \tau_t (2q-k)^\mu \frac{f_t-1}{t-q^2} \left[f_u + \hat{A} \left(1-f_u\right) \right]  \nonumber \\
             & + \tau_u (2p'-k)^\mu \frac{f_u-1}{u-p'^2} \left[f_t + \hat{A} \left(1-f_t\right) \right].   \label{eq:Cmu}
\end{align}
Here $p'$, $q$, and $k$ are the four-momenta for the outgoing $\Sigma(1385)$, outgoing $K$, and incoming photon, respectively; $\tau_{t\left(u\right)}$ denotes the isospin factor in the corresponding $t\left(u\right)$-channel hadronic vertex; $f_u$ and $f_t$ are the phenomenological form factors for $u$-channel $\Sigma(1385)$ exchange and $t$-channel $K$ exchange, respectively; $\hat{A}$ is a Lorentz-covariant, crossing-symmetric phenomenological function introduced to prevent the ``violation of scaling behavior'' as noted in Ref.~\cite{Drell:1972}. The prescription of Eq.~(\ref{eq:Cmu}) for the auxiliary current $C^\mu$ corresponds to set $\hat{h}=1-\hat{A}$ in Ref.~\cite{Wang:2020}. Following Ref.~\cite{Xiaoyun:2016}, $\hat{A}$ is taken as
\begin{equation}
\hat{A}(t,u) = A_{0}\frac{\Lambda_{c}^{4}}{\Lambda_{c}^{4}+\left(s-s_{\rm th}\right)^2},   \label{eq:A0}
\end{equation}
with
\begin{equation}
s_{\rm th}=\left(m_{\Sigma^\ast}+m_K\right)^2.
\end{equation}
Here the cutoff $\Lambda_{c}$ is fixed to be $2.5$ GeV to make $\hat{A}$ not fall off too rapidly for the energy range considered, and the strength $A_{0}$ is taken as a fit parameter.

In Ref.~\cite{Wang:2020}, we have presented the effective Lagrangians, resonance propagators, and phenomenological form factors for the $\gamma p \to K^+ \Sigma^0(1385)$ reaction. The same formulas also apply for the $\gamma n \to K^+ \Sigma^-(1385)$ reaction. For the brevity of the present paper, we do not repeat them here. For the newly introduced $u$-channel $\Sigma(1385)$ exchange, the Lagrangian for the $\gamma\Sigma^\ast\Sigma^\ast$ electromagnetic coupling reads
\begin{equation}
\mathcal{L}_{\gamma \Sigma^\ast\Sigma^\ast} = e {\bar\Sigma}^\ast_\mu A_{\alpha}\Gamma_{\gamma\Sigma^\ast}^{\alpha,\mu\nu}\Sigma^\ast_{\nu},
\end{equation}
with
\begin{align}
A_{\alpha}\Gamma_{\gamma\Sigma^\ast}^{\alpha,\mu\nu} = & \; Q_{\Sigma^\ast}A_{\alpha}\left[g^{\mu\nu}\gamma^{\alpha}- \frac{1}{2} \left(\gamma^{\mu}\gamma^{\nu}\gamma^{\alpha}+\gamma^{\alpha}\gamma^{\mu}\gamma^{\nu}\right)\right] \nonumber \\
  & - \frac{\kappa_{\Sigma^\ast}}{2M_N}\sigma^{\alpha\beta}\left(\partial_{\beta}A_{\alpha}\right)g^{\mu\nu},
\end{align}
where $Q_{\Sigma^\ast}$ is the electric charge of $\Sigma(1385)$, and $\kappa_{\Sigma^\ast}$ denotes the anomalous magnetic moment of $\Sigma(1385)$ taken as $\kappa_{\Sigma^{\ast-}}=-2.43$ from a quark model prediction \cite{Lichtenberg:1977}.

\section{Results and discussion}   \label{Sec:results}

\begin{table}[tb]
\caption{\label{Table:para} Fitted values of adjustable model parameters. The other parameters not shown in this table are taken from model III of Ref.~\cite{Wang:2020}. $R$ denotes the $\Delta(1930){5/2}^-$ resonance. The cutoff parameters $\Lambda_{K}$ and $\Lambda_{\Sigma^\ast}$ are in MeV.}
\begin{tabular*}{\columnwidth}{@{\extracolsep\fill}ccccc}
\hline\hline
$A_0$ & $\Lambda_{K}$ & $\Lambda_{\Sigma^\ast}$ & $g_{RN\gamma}^{(1)}g_{R\Sigma^\ast K}^{(1)}$ & $g_{RN\gamma}^{(2)}g_{R\Sigma^\ast K}^{(1)}$  \\
$-0.056\pm 0.009$  &  $805\pm 7$  &  $915\pm 97$  & $-8.7\pm 0.3$ & $-46.5\pm 1.8$  \\
\hline\hline
\end{tabular*}
\end{table}

\begin{figure}[htb]
\includegraphics[width=\columnwidth]{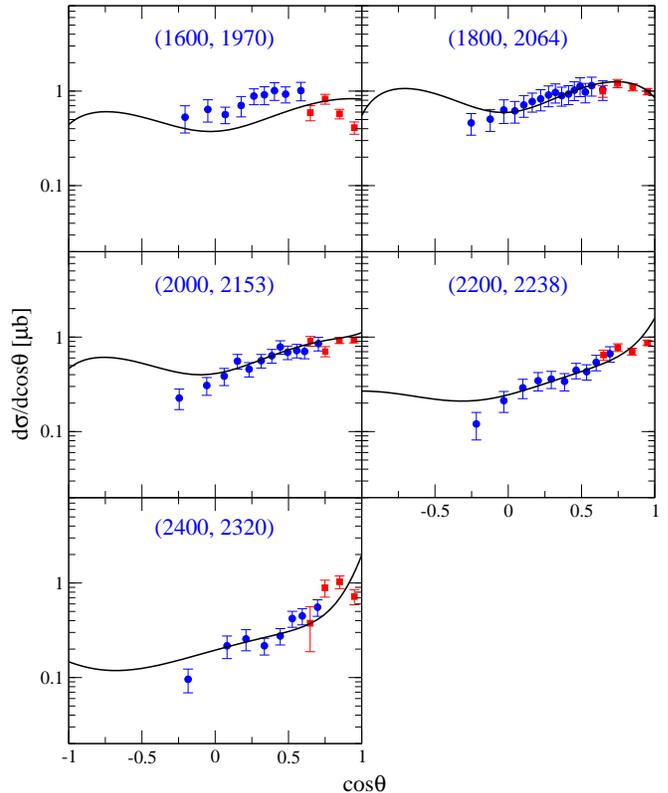}
\caption{Differential cross sections for $\gamma n \to K^+ \Sigma^-(1385)$ as a function of $\cos\theta$ in model III of Ref.~\cite{Wang:2020} with the $\Delta(1930){5/2}^-$ resonance. The red squares and blue circles denote the data from the LEPS Collaboration \cite{Hicks:2009} and CLAS Collaboration \cite{Paul:2014}, respectively. The numbers in parentheses denote the centroid value of the photon laboratory incident energy (left number) and the corresponding total center-of-mass energy of the system (right number), in MeV.}
\label{fig:dif1}
\end{figure}

\begin{figure}[htb]
\includegraphics[width=\columnwidth]{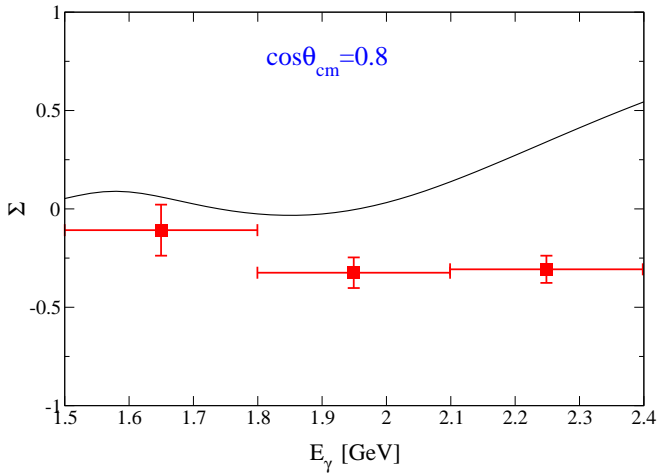}
\caption{Photo-beam asymmetries for $\gamma n \to K^+ \Sigma^-(1385)$ plotted against the photon laboratory energy $E_\gamma$ in model III of Ref.~\cite{Wang:2020} with the $\Delta(1930){5/2}^-$ resonance. The red full squares denote the data from the LEPS Collaboration \cite{Hicks:2009}. }
\label{fig:beam}
\end{figure}

\begin{figure}[htb]
\includegraphics[width=\columnwidth]{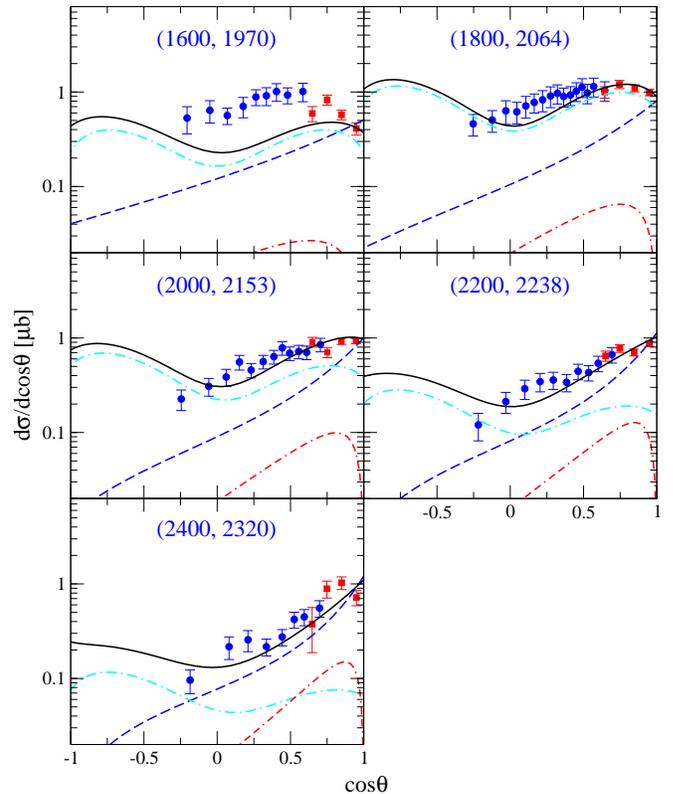}
\caption{Differential cross sections for $\gamma n \to K^+ \Sigma^-(1385)$ as a function of $\cos\theta$ (black solid lines). The red squares and blue circles denote the data from the LEPS Collaboration \cite{Hicks:2009} and CLAS Collaboration \cite{Paul:2014}, respectively. The cyan dash-dotted, blue dashed, and red dot-double-dashed lines represent the individual contributions from the $s$-channel $\Delta(1930){5/2}^-$ resonance, generalized contact term, and $t$-channel $K$ exchange, respectively. The numbers in parentheses denote the centroid value of the photon laboratory incident energy (left number) and the corresponding total center-of-mass energy of the system (right number), in MeV.}
\label{fig:dif1_2}
\end{figure}

\begin{figure}[htb]
\includegraphics[width=\columnwidth]{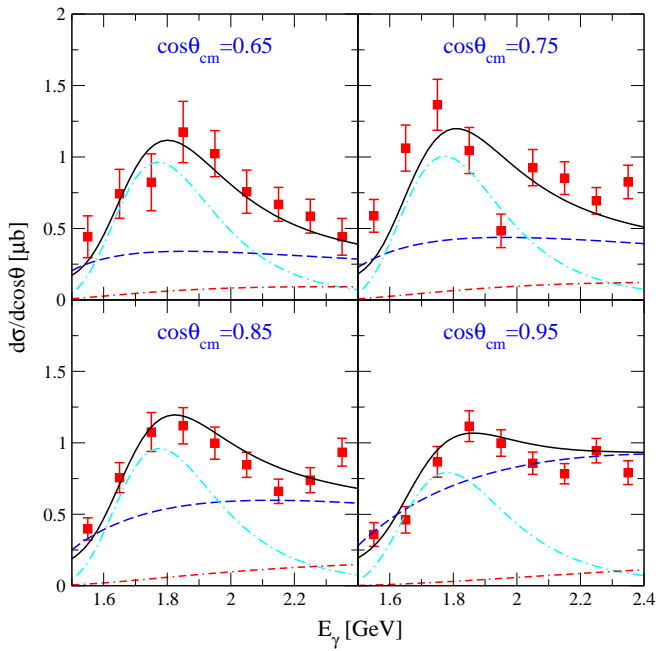}
\caption{Differential cross sections for $\gamma n \to K^+ \Sigma^-(1385)$ as a function of photon incident energy at four $\cos\theta$ intervals (black solid lines). The notations are the same as in Fig.~\ref{fig:dif1_2}. The red full squares denote the data from the LEPS Collaboration \cite{Hicks:2009}.}
\label{fig:dif2_2}
\end{figure}

\begin{figure}[htb]
\includegraphics[width=\columnwidth]{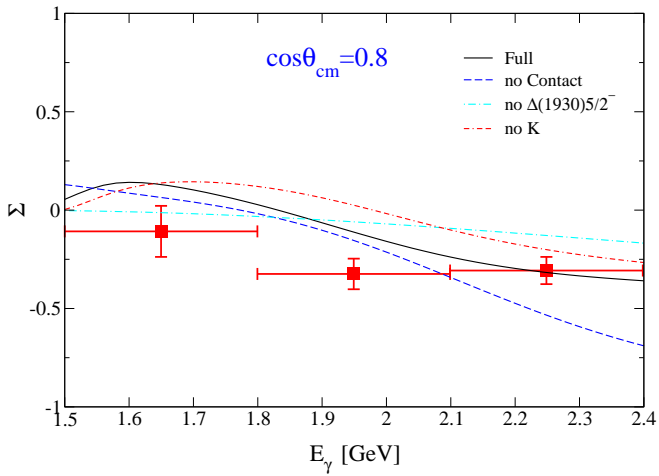}
\caption{The beam asymmetries for $\gamma n \to K^+ \Sigma^-(1385)$ plotted against photon incident energy $E_\gamma$. The red full squares denote the data from the LEPS Collaboration \cite{Hicks:2009}.}
\label{fig:beam_2}
\end{figure}

\begin{figure}[htb]
\includegraphics[width=\columnwidth]{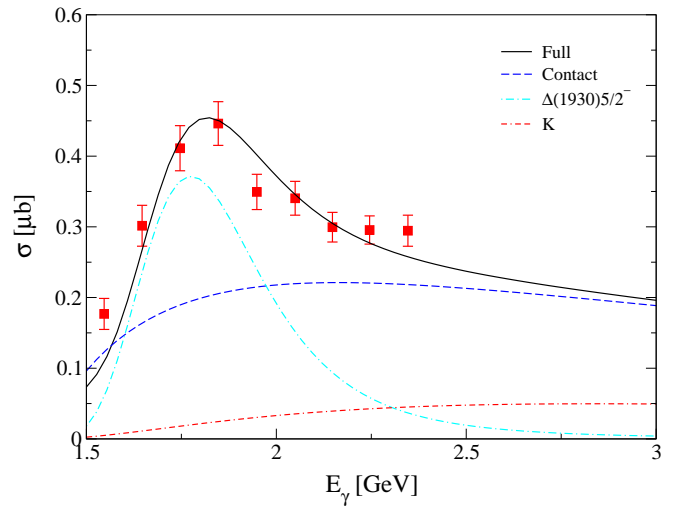}
\caption{Total cross sections for $\gamma n \to K^+ \Sigma^-(1385)$ with dominant individual contributions as a function of photon incident energy. Data (red full squares) are taken from the LEPS Collaboration \cite{Hicks:2009} but not included in the fit.}
\label{fig:total}
\end{figure}

A combined analysis of the data for both the $\gamma n\to K^+\Sigma^-(1385)$ and $\gamma p\to K^+\Sigma^0(1385)$ reactions will put more constraints on theoretical models and thus result in more reliable results. As mentioned in Sec.~\ref{Sec:intro}, in literature, the photoproduction reaction $\gamma n\to K^+\Sigma^-(1385)$ has been investigated by three theoretical works in Refs.~\cite{Zou:2010,Xiaoyun:2016,Byung:2017}. But only in a Regge model of Ref.~\cite{Byung:2017}, the differential cross-section data for $\gamma p\to K^+\Sigma^0(1385)$ were considered. There, due to the lack of resonance contributions, the cross-section data for $\gamma n\to K^+\Sigma^-(1385)$ and $\gamma p\to K^+\Sigma^0(1385)$ were only qualitatively reproduced.

In our previous work of Ref.~\cite{Wang:2020}, we have investigated the photoproduction reaction $\gamma p\to K^+\Sigma^0(1385)$ within an effective Lagrangian approach. There, apart from the $t$-channel $K$ and $K^\ast(892)$ exchanges, $s$-channel $N$ contribution, $u$-channel $\Lambda$ exchange, and generalized contact term, the contributions from a minimum number of $N$ and $\Delta$ resonances were also considered to construct the reaction amplitudes to describe the data. It was found that the high-precision differential cross-section data from the CLAS Collaboration for $\gamma p\to K^+\Sigma^0(1385)$ can be well reproduced by including the contribution from one of the $N(1895){1/2}^-$, $\Delta(1900){1/2}^-$, and $\Delta(1930){5/2}^-$ resonances in the $s$ channel.

In the present work, we plan to analyze all the available data on differential cross sections from the LEPS Collaboration and CLAS Collaboration and on photo-beam asymmetries from the LEPS Collaboration for the reaction $\gamma n\to K^+\Sigma^-(1385)$ within the same model as in our previous work of Ref.~\cite{Wang:2020} that works quite well for $\gamma p\to K^+\Sigma^0(1385)$. The purpose is to understand the reaction mechanism of $\gamma n\to K^+\Sigma^-(1385)$ and, in particular, to figure out which one of the $N(1895){1/2}^-$, $\Delta(1900){1/2}^-$, and $\Delta(1930){5/2}^-$ resonances really contributes in both the $\gamma n\to K^+\Sigma^-(1385)$ and $\gamma p\to K^+\Sigma^0(1385)$ reactions.

Most of the interaction mechanisms for $\gamma p\to K^+\Sigma^0(1385)$ and $\gamma n\to K^+\Sigma^-(1385)$ are the same except for the $u$-channel interaction and the generalized contact term. In $u$-channel amplitudes, the $\Lambda$ exchange contributes only in $\gamma p\to K^+\Sigma^0(1385)$ while the $\Sigma^-(1385)$ exchange contributes only in $\gamma n\to K^+\Sigma^-(1385)$ due to the charges of the outgoing $\Sigma(1385)$ particles. For the generalized contact term, one has different auxiliary current $C^\mu$ [c.f. Eq.~(\ref{eq:Cmu})] for $\gamma p\to K^+\Sigma^0(1385)$ and $\gamma n\to K^+\Sigma^-(1385)$ due to the fact that apart from the $t$-channel $K^+$ exchange which contributes to both the $\gamma p$ and $\gamma n$ channels, the $s$-channel $N$ exchange (longitudinal part) contributes only in the $\gamma p$ channel and the $u$-channel $\Sigma(1385)$ exchange contributes only in the $\gamma n$ channel.

In Ref.~\cite{Wang:2020}, we have reported three models for $\gamma p\to K^+\Sigma^0(1385)$, each with one of the $N(1895){1/2}^-$, $\Delta(1900){1/2}^-$, and $\Delta(1930){5/2}^-$ resonances, respectively. Here for $\gamma n\to K^+\Sigma^-(1385)$, we test these three models one by one by comparing the theoretical results from each model with the data. In all these three models, one has $A_0$ in Eq.~(\ref{eq:A0}) and the cutoff parameter $\Lambda_{\Sigma^\ast}$ for $u$-channel $\Sigma(1385)$ exchange as fit parameters, and in model I with the $N(1895){1/2}^-$ resonance, one has the resonance electromagnetic coupling constant as an additional fit parameter.

The results from these three models show that although the differential cross-section data for $\gamma n\to K^+\Sigma^-(1385)$ can be qualitatively described, the negative photo-beam asymmetries cannot be reasonably reproduced. As an illustration, we show in Fig.~\ref{fig:dif1} the differential cross sections and in Fig.~\ref{fig:beam} the photo-beam asymmetries resulted from model III of Ref.~\cite{Wang:2020} with the $\Delta(1930){5/2}^-$ resonance. One sees that the differential cross-section data are satisfactorily described, however, in the energy region of photon energy $E_\gamma>1.9$ GeV, the LEPS beam-asymmetry data show negative values while the theoretical results are positive.

We then release the resonance electromagnetic coupling constants for $\Delta$ resonances and the cutoff value for $t$-channel $K$ exchange as fit parameters instead of taking their fixed values from Ref.~\cite{Wang:2020}. By doing this, it is found that both the differential cross-section data and the photo-beam asymmetry data for $\gamma n\to K^+\Sigma^-(1385)$ can be reasonably reproduced in the model where the $\Delta(1930){5/2}^-$ resonance is included, while in other two models with inclusion of either the $N(1895){1/2}^-$ or the $\Delta(1900){1/2}^-$ resonance, the data cannot be satisfactorily described. In the following parts of the paper, we show and discuss the results from the model with the $\Delta(1930){5/2}^-$ resonance.

In Table~\ref{Table:para}, we present the fitted values of the adjustable parameters in the model with the $\Delta(1930){5/2}^-$ resonance. The values of the parameters that are not shown in this table are taken from model III of Ref.~\cite{Wang:2020}. One sees that the cutoff parameter of $t$-channel $K$ exchange is reduced from $950$ MeV in model III of Ref.~\cite{Wang:2020} to $805$ MeV. Although the $t$-channel $K$ exchange itself results in negative beam asymmetries, this diagram also contributes to the generalized contact term [c.f. Eq.~(\ref{eq:Mint})] which results in positive beam asymmetries. This explains why the beam asymmetry data for $\gamma n\to K^+\Sigma^-(1385)$ prefers a smaller cutoff value for $t$-channel $K$ exchange. The fitted value of the cutoff parameter for $u$-channel $\Sigma(1385)$ exchange has large uncertainties. This is because that we don't have data at backward angles where the $u$-channel $\Sigma(1385)$ exchange is expected to contribute significantly, and thus this parameter is not that well constrained by the available data.

The theoretical results for differential cross sections and photo-beam asymmetries are shown in Figs.~\ref{fig:dif1_2}-\ref{fig:beam_2}. There, the black solid lines represent the results from the full calculation. The cyan dash-dotted, blue dashed, and red dot-double-dashed lines represent the individual contributions from the $s$-channel $\Delta(1930){5/2}^-$ resonance, generalized contact term, and $t$-channel $K$ exchange, respectively. The contributions from other terms are too small to be clearly seen with the scale used, and thus they are not plotted. In Fig.~\ref{fig:dif1_2}, the numbers in parentheses denote the centroid value of the photon laboratory incident energy (left number) and the corresponding total center-of-mass energy of the system (right number), in MeV. One sees from Figs.~\ref{fig:dif1_2}-\ref{fig:beam_2} that all our theoretical differential cross sections and photo-beam asymmetries for $\gamma n\to K^+\Sigma^-(1385)$ agree with the corresponding data quite well, except that at the lowest energy in Fig.~\ref{fig:dif1_2}, the CLAS differential cross-section data \cite{Paul:2014} are somehow underestimated. Note that the CLAS differential cross-section data are measured at a relative large energy bin of $200$ MeV. In the first subfigure of Fig.~\ref{fig:dif1_2}, we show the theoretical results are $E_\gamma=1600$ MeV, while the CLAS data were measured at $E_\gamma = 1500-1700$ MeV. We know that near threshold the phase space is rather sensitive to the energy, which may partially explain the deviation of our theoretical differential cross sections from the data at $E_\gamma=1600$ MeV.

It is seen from Figs.~\ref{fig:dif1_2} and \ref{fig:dif2_2} that the $\Delta(1930){5/2}^-$ resonance contribution dominates the cross sections of $\gamma n\to K^+\Sigma^-(1385)$ in the relative low-energy region ($E_\gamma \leq 2000$ MeV), the contact term contribute significantly in the whole energy region considered, and the $K$ exchange makes considerable contributions at forward angles in the high energy region.

In Fig.~\ref{fig:total}, we show the total cross sections for $\gamma n\to K^+\Sigma^-(1385)$ with dominant individual contributions as a function of photon incident energy. Note that the LEPS data are measured at the angular interval $\cos\theta=0.6-1$, and correspondingly, our theoretical total cross sections are obtained by an integral of the differential cross sections at the angular interval $\cos\theta=0.6-1$. Although only the differential cross sections at forward angles are considered, consistent observations can be made from Fig.~\ref{fig:total} compared with those obtained from the differential cross sections of Figs.~\ref{fig:dif1_2} and \ref{fig:dif2_2}. The $\Delta(1930){5/2}^-$ resonance contribution dominates the cross sections below $E_\gamma = 2000$ MeV and is responsible for the bump structure exhibited by the data near the $K\Sigma(1385)$ threshold. The contact term provides significant contributions in the whole energy region considered, and actually, it occupies the main body of the background contributions. Considerable contributions can also be seen from the $t$-channel $K$ exchange at higher energies.

\section{Summary and conclusion}  \label{sec:summary}

A combined analysis of the data for $\gamma n\to K^+\Sigma^-(1385)$ and $\gamma p\to K^+\Sigma^0(1385)$ will put on more constraints on theoretical models and thus result in more reliable results. In literature, the data for the $\gamma n\to K^+\Sigma^-(1385)$ reaction have been investigated in three theoretical works \cite{Zou:2010,Xiaoyun:2016,Byung:2017}. But only in Ref.~\cite{Byung:2017}, the data for the $\gamma p\to K^+\Sigma^0(1385)$ reaction have been considered simultaneously. There, the data for both $\gamma n\to K^+\Sigma^-(1385)$ and $\gamma p\to K^+\Sigma^0(1385)$ are roughly described in a Regge model with some room left for improvements due to the lack of contributions from nucleon resonances in the Regge model.

In our previous work \cite{Wang:2020}, the high-precision differential cross section data for $\gamma p\to K^+\Sigma^0(1385)$ from the CLAS Collaboration \cite{Moriya2013} have been studied in an effective Lagrangian approach. There, we constructed the reaction amplitudes by considering the contributions from a minimum number of $N$ and $\Delta$ resonances in the $s$ channel besides the $K$ and $K^\ast(892)$ exchanges in the $t$ channel, $\Lambda$ exchange in the $u$ channel, $N$ contribution in the $s$ channel, and the generalized contact term. It was found that the data for $\gamma p\to K^+\Sigma^0(1385)$ can be well reproduced by considering one of the $N(1895){1/2}^-$, $\Delta(1900){1/2}^-$, and $\Delta(1930){5/2}^-$ resonances.

In the present work, we use the same model developed in our previous work of Ref.~\cite{Wang:2020} for the $\gamma p\to K^+\Sigma^0(1385)$ reaction to analyze the currently available differential cross-section data from the LEPS Collaboration and CLAS Collaboration and the photo-beam asymmetry data from the LEPS Collaboration for the reaction $\gamma n\to K^+\Sigma^-(1385)$, with the purpose being to understand the reaction mechanism of $\gamma n\to K^+\Sigma^-(1385)$ and, in particular, to figure out which one of the $N(1895){1/2}^-$, $\Delta(1900){1/2}^-$, and $\Delta(1930){5/2}^-$ resonances is capable for a description of the data for $\gamma n\to K^+\Sigma^-(1385)$. The interaction diagrams for these two reactions are the same except that the $u$-channel $\Lambda$ exchange in $\gamma p\to K^+\Sigma^0(1385)$ is replaced by the $\Sigma(1385)$ exchange in $\gamma n\to K^+\Sigma^-(1385)$, and the generalized contact term in these two reactions differ consequently as the $u$-channel $\Sigma(1385)$ exchange contributes only in $\gamma n\to K^+\Sigma^-(1385)$ while the $s$-channel $N$ exchange (longitudinal part) contributes only in $\gamma p\to K^+\Sigma^0(1385)$. The values of most of the model parameters are taken from Ref.~\cite{Wang:2020} with the exception of the cutoff of $t$-channel $K$ exchange and the resonance electromagnetic couplings which are adjusted to fit the data.

It was found that the available differential cross-section data the photo-beam asymmetry data for $\gamma n\to K^+\Sigma^-(1385)$ can be satisfactorily described by including the contribution from the $\Delta(1930){5/2}^-$ resonance in the $s$ channel. The $\Delta(1930){5/2}^-$ resonance contribution dominates the cross sections of $\gamma n\to K^+\Sigma^-(1385)$ below $E_\gamma = 2000$ MeV and is responsible for the near-threshold bump structure exhibited by the total cross-section data. The contact term contributes significantly in the whole energy considered, and it occupies the main body of the background contributions for $\gamma n\to K^+\Sigma^-(1385)$. The $t$-channel $K$ exchange contributes considerably at forward angles in the high energy region. The contributions from other terms are found to be rather small.

Combining the results of the present work for $\gamma n \to K^+\Sigma^-(1385)$ and of our previous work \cite{Wang:2020} for $\gamma p \to K^+\Sigma^0(1385)$, a conclusion can be made that all the available data for $\Sigma(1385)$ photoproduction on both proton and neutron targets can be well described by including the contributions from the $\Delta(1930){5/2}^-$ resonance in $s$ channel in addition to the background contributions from the $s$-, $t$-, and $u$-channel hadron exchanges and the interaction current in our effective Lagrangian approach.

Seriously speaking, the cutoff parameter for $t$-channel $K$ exchange, $\Lambda_K$, and the $\Delta(1930){5/2}^-$ resonance couplings for $\gamma p \to K^+ \Sigma^0(1385)$ should be the same as those for $\gamma n \to K^+ \Sigma^-(1385)$. However, as shown in Table I of the present work and Table II of our previous work \cite{Wang:2020}, the values of $\Lambda_K$, $g^{(1)}_{RN\gamma}g^{(1)}_{R\Sigma^\ast K}$, and $g^{(2)}_{RN\gamma}g^{(1)}_{R\Sigma^\ast K}$ from the fit of the neutron target data differ from those from the fit to the proton target data. The possible reason is that in both our previous work and the present work, the resonance hadronic coupling constant $g^{(2)}_{R\Sigma^\ast K}$ is set to be $0$ for simplicity. With this parameter left free, the model is expected to provide a self-consistent description of all the available data for both $\gamma p \to K^+\Sigma^0(1385)$ and $\gamma n \to K^+\Sigma^-(1385)$ by considering the $\Delta(1930){5/2}^-$ resonance contributions in $s$ channel in addition to the background contributions. We leave such a work to the future when more data for $\Sigma(1385)$ photoproduction become available.

\begin{acknowledgments}
This work is partially supported by the National Natural Science Foundation of China under Grants No.~12175240, No.~12147153, and No.~11635009, the Fundamental Research Funds for the Central Universities, and the China Postdoctoral Science Foundation under Grants No.~2021M693141 and No.~2021M693142. 
\end{acknowledgments}

\end{document}